\documentclass[12pt,journal,draftclsnofoot,oneside,onecolumn]{IEEEtran}

\usepackage{color,soul}
\usepackage[cmex10]{amsmath}
\usepackage{amssymb}
\usepackage{dsfont}
\usepackage{epsfig}
\usepackage{stfloats}
\usepackage{cite}
\usepackage{subfigure}

\begin{document}

\title{Cognitive Non-Orthogonal Multiple Access with Cooperative Relaying: A New Wireless Frontier for 5G Spectrum Sharing}

\author{Lu Lv, Jian Chen, Qiang Ni, Zhiguo Ding, and Hai Jiang
\thanks{Lu Lv and Jian Chen are with Xidian University; Qiang Ni and Zhiguo Ding are with Lancaster University; Hai Jiang is with University of Alberta.}
\vspace{-10mm}}
\markboth{IEEE Communications Magazine (Accepted from Open Call)}{}
\maketitle

\begin{abstract}\vspace{-3mm}
Two emerging technologies towards 5G wireless networks, namely non-orthogonal multiple access (NOMA) and cognitive radio (CR), will provide more efficient utilization of wireless spectrum in the future.  In this article, we investigate the integration of NOMA with CR into a holistic system, namely cognitive NOMA network, for more intelligent spectrum sharing. Design principles of cognitive NOMA networks are perfectly aligned to functionality requirements of 5G wireless networks, such as high spectrum efficiency, massive connectivity, low latency, and better fairness. Three different cognitive NOMA architectures are presented, including underlay NOMA networks, overlay NOMA networks, and CR-inspired NOMA networks. To address inter- and intra-network interference which largely degrade the performance of cognitive NOMA networks, cooperative relaying strategies are proposed. For each cognitive NOMA architecture, our proposed cooperative relaying strategy shows its potential to significantly lower outage probabilities. Furthermore, we discuss open challenges and future research directions on implementation of cognitive NOMA networks.
\end{abstract}\vspace{-3mm}
\IEEEpeerreviewmaketitle

\section{Introduction}
Non-orthogonal multiple access (NOMA) has been widely recognized as a promising multiple access technology to enable efficient utilization of spectrum resources in 5G wireless networks \cite{Al-Imari_Imran_2017,Octavia_JSAC2017,LDai_CM2015,Ioannis_SPL2015,Octavia_CST2017,QYang_JSAC2017}. The key idea of NOMA is to encourage spectrum sharing among multiple users within one resource block by exploiting power domain multiplexing, fundamentally differing from conventional orthogonal multiple access (OMA) technologies (which rely on time/frequency/space domain multiplexing). More recently, NOMA has been incorporated in various standardizations, for example, multi-user superposition transmission (MUST) in 3rd-generation partnership project long-term evolution (3GPP-LTE) and layered division multiplexing (LDM) in digital TV standard ATSC 3.0 \cite{Ding_JSAC2017}.

It has also been known that more efficient use of wireless spectrum can be achieved by cognitive radio (CR), where secondary users (SUs) intelligently adapt their operating parameters to access a spectrum band occupied by primary users (PUs), in an opportunistic or collaborative manner \cite{Octavia_CST2016,Goldsmith_PROC2009}. To date, exploiting CR technology to support emerging applications has received considerable attention, where novel architectures for CR networks based on full-duplex, device-to-device, and multiple-input multiple-output (MIMO) have been studied to further increase spectrum efficiency. More particularly, existing research on the combination of NOMA and CR (i.e., \cite{Lu_TVT,YLiu_TVT2016,Ding_TVT2016,WLiang_TCOM2017,Lu_TCOM}) has shown the possibility to meet 5G requirements of high throughput, massive connectivity, as well as low latency. Despite these potential benefits, building efficient cognitive NOMA is a challenging issue in practice. This is because both NOMA and CR are interference-limited, and thus,  coexistence of inter-network interference between the primary and secondary networks and intra-network interference (also called co-channel interference) caused by power domain multiplexing of NOMA undoubtedly results in severe performance degradation of reception reliability.
Therefore, it is necessary to combine NOMA with CR in an appropriate manner for minimizing the interference and better utilizing the underlying spectrum resource.

In this article, we integrate NOMA capabilities into CR concepts to achieve more intelligent spectrum sharing. We begin our attempt with a concise introduction of NOMA and CR paradigms, along with motivations of cognitive NOMA as well as cooperative strategies to further improve performance. Then we provide an in-depth overview of three cognitive NOMA architectures, namely underlay NOMA networks, overlay NOMA networks, and CR-inspired NOMA networks (termed CR-NOMA networks), and discuss their key operating principles. Note that both NOMA and CR are interference-limited, which might undermine reception reliability. To address this challenging issue, for each cognitive NOMA architecture, we propose a cooperative relaying strategy for reliability enhancement. Since studies of cognitive NOMA networks are still in a nascent stage, several potential challenges in this research field are also outlined, many of which are promising avenues for future works. Finally, we make concluding remarks to this article.

\section{Rationales of Cognitive NOMA Networks}
This section first introduces basics of NOMA and CR. Then the motivations of cognitive NOMA networks and cooperative strategies  are discussed.

\subsection{Understanding NOMA and CR}
\subsubsection{NOMA Principles}
\begin{figure}[t]
  \centering
  \includegraphics[width=0.8\textwidth]{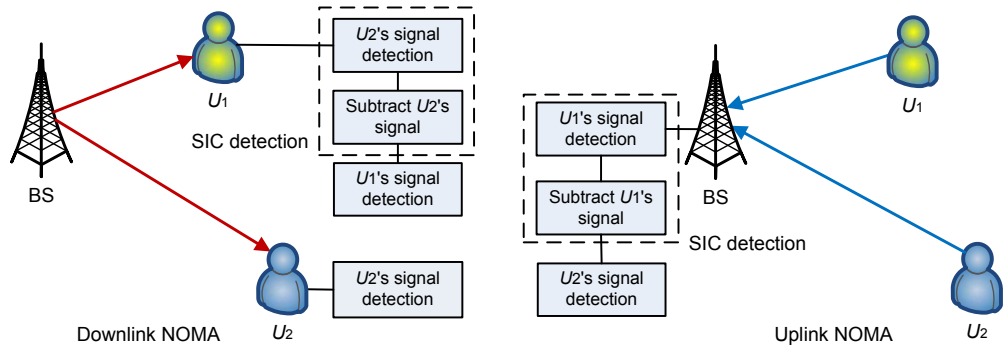}
  \caption{An illustration of NOMA systems. Left: downlink NOMA signaling; right: uplink NOMA signaling.}\vspace{-2mm}\label{NOMA-basic}
\end{figure}

The main idea of NOMA is to exploit power domain multiplexing at transmitter(s) for signal combination, and successive interference cancellation (SIC) at receiver(s) for signal detection. NOMA can be realized in downlink or uplink as follows.
\begin{itemize}
  \item {\it Downlink NOMA}: An example of downlink NOMA transmission is shown in the left-hand side of Fig.~\ref{NOMA-basic}.  Upon NOMA signaling, a base station (BS) broadcasts a signal mixture $(a_1s_1+a_2s_2)$ to user $U_1$ (who has a strong channel condition, referred to as {\it the strong user}) and $U_2$ (who has a weak channel condition, referred to as {\it the weak user}), where $s_1$ and $s_2$ are signals intended for the two users, and $a_1$ and $a_2$ are the corresponding power allocation coefficients with $a_1^2+a_2^2=1$ and $a_1<a_2$. At $U_1$, SIC is carried out to remove $s_2$ and then recover its own $s_1$. At $U_2$, its signal $s_2$ is decoded directly by treating $s_1$ as interference. Consequently, the weak user ($U_2$) suffers interference from the other user, and the strong user ($U_1$) enjoys interference-free transmission.
  \item {\it Uplink NOMA}: In uplink NOMA transmission shown in the right-hand side of Fig.~\ref{NOMA-basic}, usually a control message containing information of power allocation should be sent by the BS at the initial stage \cite{Ding_JSAC2017}. Then $U_1$ (with a strong channel condition, referred to as  {\it the strong user}) and $U_2$ (with a weak channel condition, referred to as  {\it the weak user}) transmit desired signals $s_1$ and $s_2$ to the BS using different power levels $a_1$ and $a_2$, respectively. On receiving these signals, SIC is carried out at the BS, and an optimal detection ordering would be to decode starting from the stronger signal first and moving towards the weaker signal. It is preferable to have distinct received power strength for SIC processing, and thus, we choose $a_1>a_2$ in uplink NOMA so that the BS first decodes $s_1$ and then cancels it to recover $s_2$. Therefore, the weak user ($U_2$) enjoys interference-free transmission while the strong user ($U_1$) observes interference.
\end{itemize}

\subsubsection{CR Paradigms}
A major objective of CR is to realize dynamic spectrum access/sharing by learning its surrounding environments and adapting its operating parameters. Currently, there exist three CR paradigms \cite{Goldsmith_PROC2009}
\begin{itemize}
  \item {\it Interweave}: A SU can transmit only when no PU occupies the licensed spectrum.
  \item {\it Underlay}: Concurrent primary and secondary transmissions are allowed, under the condition that the interference on the primary network is below a controllable level.
  \item {\it Overlay}: A SU provides relaying services to the primary network, and at the same time, transmits its own signal.
\end{itemize}

\subsection{What Drives Cognitive NOMA Networks?}
Both NOMA and CR target efficient spectrum utilization. They enhance spectrum efficiency from different perspectives. CR helps low-priority secondary access in an opportunistic manner (i.e., in an interweave or an underlay mode when secondary access does not affect much the primary network) or in a collaborative manner (i.e., in an overlay mode). NOMA enables multiple users to transmit simultaneously by differentiating their power levels. Thus, cognitive NOMA expects to provide more intelligent spectrum sharing, by combining the CR spectrum sharing and NOMA spectrum sharing in a constructive way to further improve the spectrum utilization.

Benefits resulted from the intelligent spectrum sharing of cognitive NOMA are listed below.
\begin{itemize}
  \item {\it Improved spectrum efficiency}: Cognitive NOMA networks can make PUs and SUs active simultaneously with acceptable reception quality. Thus, spectrum utilization efficiency is largely improved.

  \item {\it Massive connectivity}: 5G wireless networks are envisioned to support a massive number of smart devices, such as augmented reality (AR), virtual reality (VR), on-line health care, and Internet of things (IoTs). This demand can be fulfilled by cognitive NOMA networks, where multiple PUs and/or SUs can be served simultaneously in one resource block with different power levels \cite{YLiu_TVT2016}.

  \item {\it Low latency}: Transmission delay of SUs can be reduced in cognitive NOMA networks, yielding low latency performance. For example, by leveraging NOMA to underlay CR networks, multiple SUs can be connected simultaneously, fundamentally differing from the OMA scenario in which only one SU can transmit when a resource block becomes available, thus potentially reducing the transmission delay for secondary access \cite{Ding_JSAC2017}.

  \item {\it Better fairness}: Cognitive NOMA networks can guarantee improved user fairness. For instance, SUs in an underlay NOMA network have an equal chance to utilize the licensed spectrum, and a SU with a weak channel condition is allocated with more power to achieve a high data rate requirement \cite{YLiu_TVT2016}. This yields a balanced tradeoff between fairness and throughput in the secondary network.
\end{itemize}

Despite many benefits shown above, there is still room to further improve performance of cognitive NOMA. Due to the coexistence of inter- and intra-network interference in cognitive NOMA networks, as well as possible poor channel conditions of transmission links because of severe path loss and/or deep fading, outage performance in cognitive NOMA networks may be degraded considerably. To tackle this challenge, we further propose to use cooperative relaying in cognitive NOMA networks, which shows a great potential to improve reception reliability.

\section{Cognitive NOMA Architectures and Cooperative Relaying Strategies}
In this section, we elaborate on existing cognitive NOMA architectures, including underlay NOMA networks, overlay NOMA networks, and CR-NOMA networks. We also show how cooperative relaying can help improve reception quality in each cognitive NOMA architecture.

\subsection{Underlay NOMA Networks}
\begin{figure}[t]
  \centering
  \includegraphics[width=0.8\textwidth]{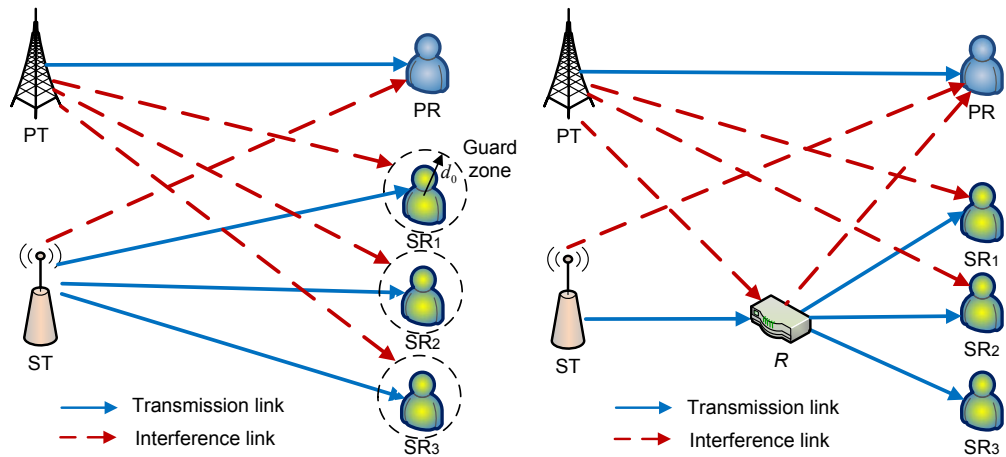}\vspace{-3mm}
  \caption{Underlay NOMA architectures. Left: non-cooperative underlay NOMA network; right: cooperative underlay NOMA network.}\vspace{-3mm}\label{underlay-NOMA}
\end{figure}

An example of underlay NOMA networks is depicted in the left-hand side of Fig.~\ref{underlay-NOMA}, where a secondary transmitter (ST) serves multiple secondary receivers (SRs) directly by NOMA signaling, given that the interference inflicted at the primary receiver (PR) is controllable. Compared with underlay OMA networks,  more efficient spectrum utilization can be achieved. This is due to the fact that, by employing NOMA in a secondary network, more SRs can simultaneously receive individual signals in the same shared spectrum, which improves connectivity of SRs and achieves high throughput for secondary access. On the other hand, in underlay OMA networks, only one SR is allowed for transmission and other SRs should wait until the current transmission is completed (i.e., the SRs utilize the spectrum in a sequential manner).

Compared with traditional NOMA networks, underlay NOMA networks face challenges of more strict interference management, as follows.
\begin{itemize}
\item NOMA assisted SRs suffer not only intra-network interference within the secondary network, but also inter-network interference generated from primary transmitters (PTs). Therefore, signal detection at NOMA assisted SRs is largely affected in presence of strong interference. In this regard, an interference guard zone aided scheme in underlay NOMA has been proposed to improve reception reliability \cite{YLiu_TVT2016}. More specifically, each NOMA assisted SR is protected by a circle with a radius of $d_0$ (called a guard zone, as shown in Fig.~\ref{underlay-NOMA}), and in the circle there is no PT. It has been shown that acceptable outage performance is achieved for secondary transmissions, and full diversity order is obtained at each NOMA assisted SR.

\item As a distinct feature of the underlay CR paradigm, transmissions of the primary network should not be interfered with when allowing secondary access. This means that power control at STs is critical to underlay NOMA networks, in the sense that the inter-network interference at a PR should be lower than a preset threshold. Moreover, power allocation at STs also needs to ensure that the intra-network interference among NOMA assisted SRs is carefully regulated, such that diverse quality-of-service (QoS) requirements of SRs can be strictly satisfied.
\end{itemize}

Traditional underlay CR is often limited to short-range communications, due to restricted transmit power levels at STs \cite{Octavia_CST2016}. In underlay NOMA, radio coverage of a secondary network is more strictly bounded, due to coexistence of inter- and intra-network interference. Note also that it is not always possible to find an appropriate interference guard zone for each SR, since the distance from PTs to SRs may be short in practical CR scenarios and such a guard zone may not exist. To address these challenges, we propose a cooperative underlay NOMA architecture by enlisting the help of a relay, to provide radio coverage extension and outage minimization for underlay NOMA networks. The proposed cooperative underlay NOMA network is depicted in the right-hand side of Fig.~\ref{underlay-NOMA}, where multiple NOMA assisted SRs are connected with a ST via an amplify-and-forward (AF) relay $R$. In this scenario, $R$ amplifies power superimposed signals sent by the ST using an amplifying coefficient. Thereby, the achievable signal-to-interference-plus-noise ratios (SINRs) at NOMA assisted SRs can be increased significantly, which is beneficial for SIC processing. In turn, radio coverage of the secondary network can be potentially extended without degrading its outage performance. Furthermore, the transmit power of the ST and $R$ can be lowered thanks to the small path-loss by cooperative relaying, which better ensures the interference constraint at the PR. Secondary outage performance of the cooperative/non-cooperative underlay NOMA architectures are shown in Fig.~\ref{sim-overlay}, in which NOMA user grouping includes all SRs (note that other grouping methods \cite{Ding_TVT2016} can also be used). It is clear that the proposed cooperative NOMA architecture can significantly improve secondary outage performance. It can also be seen that, for a specific outage probability, the cooperative NOMA needs smaller signal-to-noise ratio (SNR), which also means less transmit power and less energy consumption, than its non-cooperative counterpart.

\subsection{Overlay NOMA Networks}
\begin{figure}[t]
  \centering
  \includegraphics[width=0.9\textwidth]{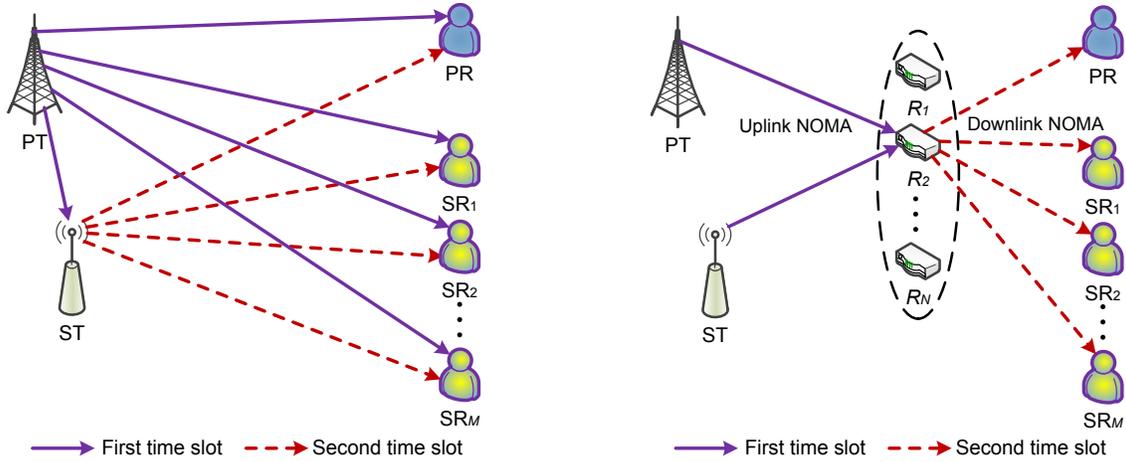}
  \caption{Overlay NOMA architectures. Left: overlay NOMA network; right: cooperative overlay NOMA network.}\vspace{-2mm}\label{overlay-NOMA}
\end{figure}

An illustration of overlay NOMA networks is depicted in the left-hand side of Fig.~\ref{overlay-NOMA}, where a ST helps forward a PT's signal to a PR, and simultaneously sends its own signals to multiple SRs, by using NOMA principles. The NOMA enabled spectrum sharing protocol is described as follows \cite{Lu_TVT}.
\begin{itemize}
  \item In the first time slot, the PT sends its primary signal to the PR. The signal is also received by the ST and the SRs.
  \item In the second time slot, the ST regenerates the primary signal and superimposes with its own signals by using NOMA. Then ST sends the signal mixture to the PR and SRs.
\end{itemize}
Compared with traditional NOMA, overlay NOMA needs extra efforts in handling inter-network interference, as follows. Since the PR receives the primary signal in both time slots, it treats secondary signals as noise and decodes the primary signal by using maximal ratio combining (MRC). At a SR, it first decodes the primary signal by using MRC, and then employs SIC to sequentially decode secondary signals until its own signal is retrieved. Several benefits can be obtained from this overlay NOMA network, such that (i) superior spectrum utilization: the underlying spectrum is exploited by simultaneous primary and secondary transmissions, giving efficient two-time-slot communications, and (ii) enhanced reception performance: inter-network interference from the primary network to the secondary network is cancelled at SRs, and primary outage can be largely improved thanks to an increased diversity gain by ST relaying. Therefore, a balanced tradeoff between spectrum efficiency and reception reliability can be achieved.

\begin{figure}[t]
\centering
\includegraphics[width=0.6\textwidth]{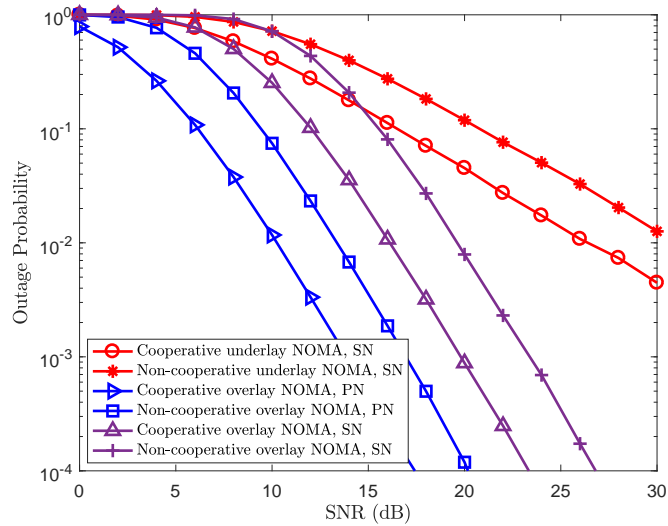}\\\vspace{-5mm}
 \caption{Outage performance of cooperative/non-cooperative NOMA schemes for underlay and overlay paradigms in single-input single-output (SISO) scenario, where ``PN'' is short for ``primary network'' and ``SN'' is short for ``secondary network''. All channels experience independent and identically distributed (i.i.d.) Rayleigh fading. For both paradigms, we consider one PR and two NOMA assisted SRs. The target data rate for PR is $0.8$ bps/Hz, and the target data rate for each SR is $0.5$ bps/Hz. For underlay paradigm, the average channel gains from ST to SRs and $R$ are 1 and 3, and average channel gain from $R$ to SRs is 3. The interference from PT is Gaussian noise with power $10$ dB \cite{YLiu_TVT2016}. The power allocation coefficients for the two SRs are 0.8 and 0.2, respectively. For overlay paradigm, the average channel gains are set the same as 2, and the number of relays is $N=3$. The power allocation coefficients for the PR and the two SRs are set to 0.8, 0.15, and 0.05, respectively}\vspace{-2mm}\label{sim-overlay}
\end{figure}

When the primary and secondary networks experience deep fading or shadowing, the PT$\rightarrow$PR and ST$\rightarrow$SR direct links may have poor channel conditions. To address this problem, we propose to implement a cooperative relay to improve outage performance. The proposed cooperative overlay NOMA network is shown in the right-hand side of Fig.~\ref{overlay-NOMA}. One advantage of the proposed framework is that both PT and ST can share the same relay. In particular, transmissions over the two time slots are as follows.
\begin{itemize}
  \item In the first time slot, uplink NOMA is used at the PT and ST where a large power allocation coefficient is assigned to the primary signal due to its high priority \cite{Octavia_CST2016}. With SIC, the relay performs decoding of the primary and secondary signals sequentially.
  \item Based on its decoding result in the first time slot, the relay dynamically selects a proper multiple access mode in the second time slot: (i) if both primary and secondary signals are decoded correctly, downlink NOMA begins immediately, and (ii) if only the primary signal is retrieved, the relay solely forwards the primary signal by conventional OMA.
\end{itemize}
It is clear that reception reliability of both PR and SRs can be significantly improved by using the relay sharing cooperative overlay NOMA architecture. There are other promising directions in the field of relay sharing cooperative overlay NOMA networks. For example, when multiple relays are available, spatial diversity gains offered by relay selection and/or collaborative relay beamforming can be exploited to further improve the outage performance. Figure~\ref{sim-overlay} (the figure that we used to show performance of cooperative underlay NOMA) illustrates the primary and secondary outage performance of the cooperative/non-cooperative overlay NOMA schemes. Here NOMA user grouping includes the PR (with large power allocation coefficient) and all SRs (with small power allocation coefficients). It is demonstrated that the cooperative overlay NOMA outperforms its non-cooperative counterpart in terms of more reliable primary and secondary transmissions.

\subsection{CR-NOMA Networks}
\begin{figure}[t]
  \centering
  \includegraphics[width=0.8\textwidth]{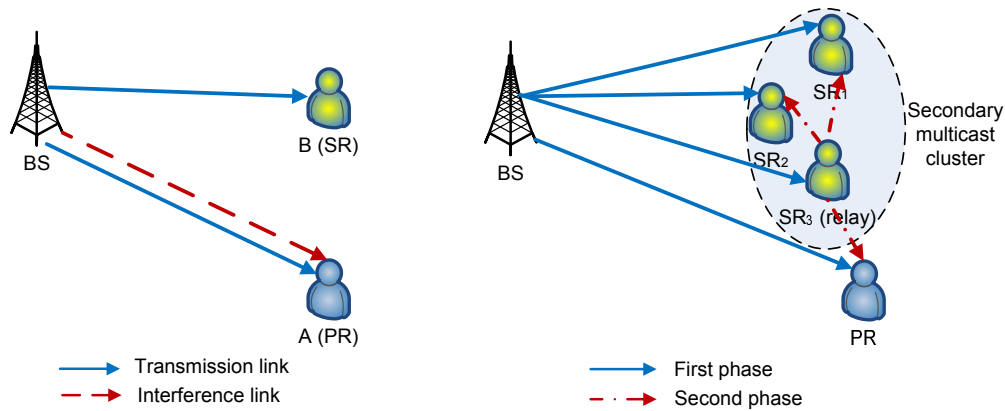}
  \caption{CR-NOMA architectures. Left: simple CR-NOMA network; right: cooperative CR-NOMA network.}\vspace{-2mm}\label{CR-NOMA}
\end{figure}

A simple CR-NOMA network  is shown in the left-hand side of Fig.~\ref{CR-NOMA}, in which a time slot is assigned to user A for downlink transmission. User A's channel is weak. If OMA is applied, user A solely occupies the channel, and thus, the spectrum is not efficiently utilized due to the weak channel. To improve spectrum efficiency, NOMA can be applied to user A and another user, say user B, that has a strong channel. This actually fits within the CR concept: user A is assigned the time slot, and thus, is the PR; user B is not assigned the time slot, but accesses the spectrum in the slot, and thus, is a SR.\footnote{When user A's channel is strong, NOMA is not applied because in this scenario, NOMA does not bring many benefits in spectrum efficiency compared to OMA.} Accordingly, this setting is referred to as CR-inspired NOMA or CR-NOMA \cite{Ding_TVT2016}. A QoS-guaranteed power allocation scheme, which divides the power into two parts for the PR's reliable reception and the SR's opportunistic transmission \cite{Ding_TVT2016}, can be applied to increase network throughput and promise user fairness. Spectrum efficiency is largely improved since the SR has a strong channel condition. In \cite{WLiang_TCOM2017}, user pairing for CR-NOMA networks with multiple pairs of PRs and SRs is investigated, where a low-complexity matching algorithm is designed to increase individual data rates of the paired users and the throughput of primary and secondary networks.

\begin{figure}[t]
\centering
\includegraphics[width=0.6\textwidth]{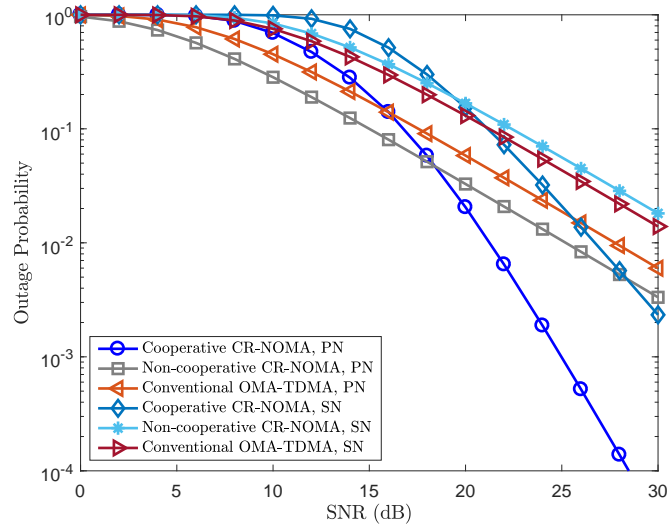}\\\vspace{-5mm}
\caption{Outage performance of cooperative/non-cooperative CR-NOMA schemes and an OMA-TDMA scheme in SISO scenario. All channels are i.i.d. Rayleigh fading. For illustration purpose, the average channel gains from the BS to the PR, from the BS to SRs, and from SRs to the PR are set to 0.5, 1 and 1, respectively, and the average channel gains among all SRs are set to 2. The target data rate for PR is $1$ bps/Hz, and the target data rate for SRs is $1.5$ bps/Hz. The power allocation coefficients for PR and SRs are 0.8 and 0.2, respectively.}\vspace{-2mm}\label{sim-CR-NOMA}
\end{figure}

In CR-NOMA networks, the SR enjoys interference-free transmission, while the PR suffers from inter-network interference (i.e., interference from the secondary transmission). Thus, the received SINR at the PR in CR-NOMA is smaller than that in the case when the PR solely occupies the spectrum. Furthermore, if the interference from the simultaneous secondary transmission is strong, primary outage may happen. Therefore, the PR may not have an incentive to share the spectrum with the SR. To overcome this challenge, a SR can be recruited as a cooperative relay to help improve reception reliability of the PR. This cooperation is particularly preferred when we need to send multicast content to SRs, as multiple SRs will provide additional benefits by appropriate user scheduling strategies. The proposed cooperative CR-NOMA is shown in the right-hand side of Fig.~\ref{CR-NOMA}, with two-phase transmissions. In the first phase, the BS sends a unicast signal to the PR and a multicast signal to the SRs. A SR that decodes both unicast and multicast signals successfully can be designated as a relay, and superimposes these signals for transmission to the PR and other SRs in the second phase \cite{Lu_TCOM}. In each phase, a large power allocation coefficient is assigned to the unicast primary signal, and a small power allocation coefficient is assigned to the multicast secondary signal. As a result, outage performance of the primary and secondary networks is boosted significantly. The performance of the cooperative CR-NOMA architecture can be further improved by using a dynamic multiple access scheme which switches between cooperative NOMA and cooperative OMA \cite{Lu_TCOM}. Figure~\ref{sim-CR-NOMA} shows the outage performance improvement of the cooperative CR-NOMA network with one PR and two SRs. It is clear that for both primary and secondary networks, the outage probability curves decrease faster in the cooperative CR-NOMA than those of non-cooperative CR-NOMA and OMA-time division multiple access (TDMA). Thus, enhanced primary and secondary outage performance is achieved by cooperative CR-NOMA.

\section{Open Challenges and Future Research Directions}
In this section, for implementing cognitive NOMA networks, we discuss some potential challenges and future research directions, most of which are also highly relevant to cooperative cognitive NOMA.

\subsubsection{Interference Management}
Cognitive NOMA networks retain highly interference-limited. Consider underlay NOMA networks, SRs suffer not only intra-network interference caused by power domain multiplexing, but also inter-network interference due to primary transmission. Moreover, total interference observed at a PR should be constrained by a controllable level. Therefore, interference management plays an important role in the design of cognitive NOMA networks. Solutions well known for conventional wireless networks, such as interference alignment and joint transceiver beamforming, can be applied to mitigate inter-network interference. In addition, power allocation should be carefully designed to minimize the negative impact of intra-network interference to underlay NOMA networks.

\subsubsection{Imperfect Channel State Information (CSI)}
Most existing cognitive NOMA research assumes that perfect CSI is available. However, in practice, channel estimation errors, mobility and feedback delay render the case of imperfect CSI, which potentially degrades the system performance. In underlay NOMA networks, imperfect CSI results in an inappropriate power allocation at the ST, which further leads to additional interference at the PR and error propagation at SRs. Furthermore, in cooperative overlay NOMA and CR-NOMA networks, a relay may be wrongly selected with a delayed version of CSI, therefore deteriorating the outage performance due to a diversity order loss for both networks. To this end, novel transmission designs which are robust to CSI errors should be developed for cognitive NOMA networks.

\subsubsection{Energy Efficiency}
Future wireless networks are expected to be green with very low energy consumption \cite{Zhang_Wang_TVT2017}. Several recent works have studied spectrum efficiency and reliability in cognitive NOMA, while the problem of energy efficiency maximization in cognitive NOMA networks is still unexplored, and can be investigated in future research. Moreover, simultaneous wireless information and power transfer can be introduced to overlay NOMA networks, where the ST extracts both information and energy from the PT's signals, and then uses the harvested energy to simultaneously serve the PR and SRs using NOMA signaling, thus achieving joint spectrum and energy efficiency. In this context, spectrum and energy efficiency tradeoff is an important metric to evaluate the overall network performance. Therefore, designs of efficient resource allocation algorithms are more than necessary.

\subsubsection{Multi-Carrier Cognitive NOMA}
Interweave NOMA is another promising paradigm for cognitive NOMA networks, which can be regarded as a multi-carrier cognitive NOMA architecture. Specifically, all SUs are first divided into multiple groups. Then the SUs in each group are served with NOMA signaling in the same resource block which is detected available via spectrum sensing, and different groups are allocated to different orthogonal resource blocks for communications. Developing corresponding resource allocation algorithms plays a crucial role in improving network performance and user fairness. However, this seems rather challenging as problems of spectrum sensing, user grouping, and subcarrier/power allocation are coupled together, which deserves more research efforts.

\subsubsection{Cognitive MIMO-NOMA}
The application of MIMO to NOMA is capable of providing additional performance gains as spatial diversity is exploited \cite{Octavia_JSAC2017}, and the extension to cognitive MIMO-NOMA is expected to still preserve this benefit. Nevertheless, designing efficient cognitive MIMO-NOMA networks is a challenging task. For example, in downlink underlay MIMO-NOMA networks, user ordering is difficult, because 1) channels of the users are represented by matrices or vectors (different from the single-input single-output [SISO] case in which the channels are represented by scalars), and 2) the users may experience different inter-network interference, and the interference at each user is also represented by a matrix or a vector. Power allocation is an opening topic, in order to restrict the resultant interference to the primary network and improve the sum rate of the secondary network. Joint transceiver beamforming is a method to achieve this, at a price of increased complexity.

\subsubsection{Relay Selection/User Scheduling}
As aforementioned, deployment of cooperative relaying has shown its great potential to improve reception reliability in cognitive NOMA networks. When multiple relays and/or SUs are available, relay selection/user scheduling is an effective yet simple approach to exploit multiuser diversity. However, conventional relay selection/user scheduling strategies mainly focus on performance of a single receiver, and thus, cannot be directly applied to cognitive NOMA networks since reception reliability of multiple NOMA assisted PUs and/or SUs should be jointly guaranteed. This motivates the design of advanced relay selection/user scheduling for future cognitive NOMA networks.

\subsubsection{Physical Layer Security}
Physical layer security is a challenging topic for cognitive NOMA networks. As previously stated, cognitive NOMA networks are vulnerable to interference. This, however, is a critical weakness that can be exploited by a denial-of-service attacker by emitting harmful radio signals to interfere with SIC processing at NOMA assisted SUs. Another security issue for cognitive NOMA networks is that a relay used for cooperation may be compromised. In other words, the untrusted relay intends to eavesdrop the confidential information for its own purpose. The application of physical layer security technologies (i.e., cooperative jamming) is a promising solution to  prevent information leakage.

\subsubsection{Full-Duplex}
Integration of full-duplex technology to cognitive NOMA can further enhance spectrum efficiency. With a full-duplex relay, there is no need for an extra time slot for relaying. When the transmitter(s) and receiver(s) are with full-duplex capability, downlink and uplink transmissions can be carried out over the the same spectrum simultaneously. However, the benefit comes with some costs. Consider underlay NOMA with full duplex as an example, in which a secondary BS has full-duplex NOMA communications with multiple SUs over the same spectrum. For downlink reception, each SU experiences inter-network interference, intra-network interference, its residual self-interference, as well as interference from other SUs' uplink transmissions. An interesting research topic is on analytical evaluation of benefits and costs of full-duplex in cognitive NOMA. It is also crucial to investigate optimal resource allocation in full-duplex cognitive NOMA, which should jointly consider user pairing, power allocation, and interference management.

\section{Conclusion}
In this article, rationales of cognitive NOMA networks have been first illustrated. Then the state-of-the-art cognitive NOMA architectures, including underlay NOMA networks, overlay NOMA networks, and CR-NOMA networks, have been discussed in details. Cooperative relaying strategies in cognitive NOMA networks have also been proposed to improve reception reliability, with cost of installing relays. Some open challenges and future research trends in the context of cognitive NOMA networks have been discussed as well.

\section*{Acknowledgment}
The work of L. Lv and J. Chen was supported in part by the National Natural Science Foundation of China under Grants 61771366 and 61601347, and in part by the ``111'' project of China under Grant B08038. The work of Q. Ni was supported by the EU FP7 CROWN project under Grant PIRSES-GA-2013-610524. The work of Z. Ding was supported in part by the UK EPSRC under Grant EP/N005597/1, and in part by H2020-MSCA-RISE-2015 under Grant 690750. The work of H. Jiang was supported by the Natural Sciences and Engineering Research Council of Canada.

\newpage
\begin{IEEEbiographynophoto}{Lu Lv}
[S'17] (lulv@stu.xidian.edu.cn) is currently working toward the Ph.D. degree of Communication and Information Systems in Xidian University, China. Since November 2016, he has been with the Department of Electrical and Computer Engineering, University of Alberta, Canada, as a visiting Ph.D. student. His research interests include cooperative relaying, non-orthogonal multiple access, and physical layer security.
\end{IEEEbiographynophoto}

\begin{IEEEbiographynophoto}{Jian Chen}
[M'14] (jianchen@mail.xidian.edu.cn) received the B.Sc. degree from Xi'an Jiaotong University in 1989, the M.Sc. degree from Xi'an Institute of Optics and Precision Mechanics of Chinese Academy of Sciences in 1992, and the Ph.D. degree from Xidian University in 2005. He is currently a professor at the School of Telecommunications Engineering, Xidian University, China. His research interests include cognitive radio, wireless sensor networks, and heterogeneous networks.
\end{IEEEbiographynophoto}

\begin{IEEEbiographynophoto}{Qiang Ni}
[M'04, SM'08] (q.ni@lancaster.ac.uk) received the Ph.D. degree from Huazhong University of Science and Technology, China. He is a professor of communications and networking and the head of the Communication Systems Group at the School of Computing and Communications, Lancaster University, UK. His main research interests are wireless communications and networking. He was an IEEE 802.11 Wireless Standard Working Group voting member and a contributor to IEEE wireless standards.
\end{IEEEbiographynophoto}

\begin{IEEEbiographynophoto}{Zhiguo Ding}
[S'03, M'05, SM'15] (z.ding@lancaster.ac.uk) received the Ph.D. degree from Imperial College London in 2005, and is currently a chair professor at Lancaster University, UK. His research interests include 5G communications, MIMO and relaying networks, and energy harvesting. He serves as an Editor for several journals including {\it IEEE Transactions on Communications}, {\it IEEE Transactions on Vehicular Technology}, and {\it Wireless Communications and Mobile Computing}.
\end{IEEEbiographynophoto}

\begin{IEEEbiographynophoto}{Hai Jiang}
[M'07, SM'15] (hai1@ualberta.ca) received the B.Sc. and M.Sc. degrees in electronics engineering from Peking University, Beijing, China, in 1995 and 1998, respectively, and the Ph.D. degree in electrical engineering from the University of Waterloo, Waterloo, Ontario, Canada, in 2006. He is currently a professor at the Department of Electrical and Computer Engineering, University of Alberta, Canada. His research interests include radio resource management, cognitive radio networking, and cooperative communications.
\end{IEEEbiographynophoto}

\end{document}